\documentclass[prd,showpacs,showkeys,nofootinbib,floatfix,%%eqsecnum,
               fleqn,preprint,12pt,tightenlines]{revtex4-1} %for preprint

\usepackage{amsmath,amssymb,revsymb,graphicx,dcolumn}

\newcommand{\version}{v4} %%true version v3.992
\newcommand{\beq}{\begin{equation}}
\newcommand{\eeq}{\end{equation}}
\newcommand{\beqa}{\begin{eqnarray}}
\newcommand{\eeqa}{\end{eqnarray}}
\newcommand{\bsubeqs}{\begin{subequations}}
\newcommand{\esubeqs}{\end{subequations}}

\def\id{\makebox[0.6ex][l]{$1$}{\rm l}}   %%frk from KlinkhamerSchimmel2002

\begin{document}

\begin{widetext}
\noindent arXiv:2008.01058 \hfill KA--TP--09--2020\;(\version)
%
%%\noindent journal  \hfill    arXiv:2008.01058\;(\version)
%
%\noindent \hfill KA--TP--09--2020\;(\version)
%
\newline\vspace*{3mm}
\end{widetext}

\title{IIB matrix model: Extracting the spacetime points}%
\vspace*{3mm}

\author{F.R. Klinkhamer}
\email{frans.klinkhamer@kit.edu}
\affiliation{Institute for Theoretical Physics,
Karlsruhe Institute of Technology (KIT),\\
76128 Karlsruhe, Germany\\}

\begin{abstract}
\vspace*{5mm}\noindent
Assuming that the large-$N$ master field
of the Lorentzian IIB matrix model
has been obtained, we go through the procedure of how
the coordinates of emerging spacetime points can be extracted.
Explicit calculations with test master fields
suggest that the genuine IIB-matrix-model master field may have
a fine-structure that is essential for producing the
spacetime points of an expanding universe.
\end{abstract}

%\pacs{04.20.Cv, 98.80.Bp, 98.80.Jk}
%\keywords{general relativity, big bang theory,
%          mathematical and relativistic aspects of cosmology}

%%98.80.Bp Origin and formation of the Universe
%%11.25.-w Strings and branes
%%11.25.Yb M theory


\maketitle

%%\newpage%%tmp
\section{Introduction}
\label{sec:Intro}

The IIB matrix model~\cite{IKKT-1997,Aoki-etal-review-1999}
has been studied numerically in its Lorentzian
version~\cite{KimNishimuraTsuchiya2012,NishimuraTsuchiya2019,%
Hatakeyama-etal2020}.
But how, conceptually, a classical spacetime emerges
in Refs.~\cite{IKKT-1997,Aoki-etal-review-1999,%
KimNishimuraTsuchiya2012,NishimuraTsuchiya2019,Hatakeyama-etal2020}
was unclear.

It has been suggested, 
%in App.~B of Ref.~\cite{Klinkhamer2019-emergence}, 
in App.~B of the preprint version~\cite{Klinkhamer2019-emergence-v6} %%XXX
of Ref.~\cite{Klinkhamer2019-emergence-PTEP},
that the large-$N$ master field must play a crucial role for the
emergence of a classical spacetime.
As a follow-up, Ref.~\cite{Klinkhamer2020} presented
an explicit (coarse-graining) procedure
for extracting classical spacetime from the master field.

Here, we give some numerical results to illustrate the
procedure of Ref.~\cite{Klinkhamer2020}, as regards the
extraction of the spacetime points
(the extraction of a spacetime metric is more difficult
and will not be discussed here).
First, we consider a test master field
with randomized entries on a band diagonal
and, then, we consider a specially designed test master field
with a deterministic fine-structure, which gives rise
to multiple strands of spacetime
that appear to fill out an expanding universe.
This last type of test master field provides
an \emph{existence proof} that there can be master fields
for which the procedure of Ref.~\cite{Klinkhamer2020}
produces more or less acceptable spacetime points.

%%\newpage%%tmp
\section{Test master fields}
\label{sec:Test-master-fields}

The discussion of the temporal test matrix is relatively simple,
as the master field $\underline{\widehat{A}}^{\;0}$ is assumed to
have been diagonalized and ordered
by an appropriate global gauge transformation~\cite{Klinkhamer2020}.
This $N\times N$ traceless Hermitian test matrix is obtained as follows:%
\bsubeqs\label{eq:Ahat0-test}
\beqa
\underline{\widehat{A}}^{\;0}_\text{\,test}
&=& \text{diag}\,
\Big[\overline{\alpha}(1), \,\ldots\, ,\,\overline{\alpha}(N)\Big]\,,
\\[2mm]
\overline{\alpha}(i)
&=&
\widetilde{\alpha}(i)
-\frac{1}{N}\;\left(\sum_{j=1}^{N}\widetilde{\alpha}(j)\right)\,,
\\[2mm]
\label{eq:Ahat0-test-alphatilde}
\widetilde{\alpha}(i)
&=& \text{rand}\left[\frac{i-1}{N},\,\frac{i}{N}\right]\,,
\eeqa
\esubeqs
with $i \in \{1,\,\ldots ,\,  N\}$ and $\text{rand}\big[x,\,y\big]$
giving a uniform pseudorandom real number in the interval $[x,\,y]$.

Next, we construct a test master field
$\underline{\widehat{A}}^{\;1}_\text{\,test-1}$
with a band-diagonal structure of width $\Delta N$ and
average absolute values along the diagonal given by
a parabola with approximate value $1$ halfway ($i \sim N/2$)
and approximate value $2$ at the edges ($i=1$ and $i=N$).
Specifically, this $N\times N$ traceless Hermitian matrix with
bandwidth $\Delta N$ (assumed to be even) is obtained as follows:%
\bsubeqs\label{eq:Ahat1-test-1}
\beqa
\label{eq:Ahat1-test-1-Hermiteanized}
\underline{\widehat{A}}^{\;1}_\text{\,test-1}
&=&
\frac{1}{2}\;
\left(\overline{A}^{\;1}+\overline{A}^{\,1\;\dagger}\right)
-\frac{1}{N}\;\text{tr}\,
\left[\frac{1}{2}\;\left(\overline{A}^{\;1}
      +\overline{A}^{\,1\;\dagger}\right)\right]\,
\id_{\,N}\,,
\eeqa
%%\\[2mm]
\beqa
\label{eq:Ahat1-test-1-randomized-band}
\left(\overline{A}^{\;1} \right)_{i,\,j}
&=&
\begin{cases}
 r_{\chi}\;\overline{x}^{\;1}(i) \,,   &
 \text{for}\;  j-\Delta N/2 \leq i \leq j+\Delta N/2\,,
 \\[2mm]
 0 \,,   &  \text{otherwise} \,,
\end{cases}\,,
\eeqa
%%\\[2mm]
\beqa
\label{eq:Ahat1-test-1-range}
r_{\chi}&=&  (1-\chi)\;\text{rand}\big[-2,\,+2\big]
             + \chi\;\text{rand}\big[\pm 1\big]\,,
\eeqa
%%\\[2mm]
\beqa
\label{eq:Ahat1-test-1-x1bar}
\overline{x}^{\;1}(i)&=&
1 + \left[\left(i-\frac{1}{2}\right)\frac{2}{N} -1 \right]^2\,,
\eeqa
%%\\[2mm]
\beqa
\label{eq:Ahat1-test-1-chi}
\chi & \in& \{0,\,1\}\,,
\eeqa
\esubeqs
with $\id_{\,N}$ the $N\times N$ identity matrix, indices
$i$ and $j$ taking values in $\{1,\,\ldots ,\,  N\}$,
$\text{rand}\big[x,\,y\big]$
defined below \eqref{eq:Ahat0-test-alphatilde},
and  $\text{rand}\big[\pm 1\big]$
giving $+1$ with probability $1/2$ and $-1$ with probability $1/2$..
The parameter $\chi$ distinguishes between a
continuous or a discrete range for the
randomized entries on the individual rows
of the band diagonal of the matrix.
The ``expanding'' behavior \eqref{eq:Ahat1-test-1-x1bar},
with a minimum at the halfway point,
mimics the numerical results obtained in
Refs.~\cite{KimNishimuraTsuchiya2012,NishimuraTsuchiya2019,Hatakeyama-etal2020},
assuming that $\overline{x}^{\;1}$ corresponds to one of the
``large'' dimensions of the $3+6$ split.
The numerical results of these last references
may, in fact, give a rough approximation of the genuine
IIB-matrix-model master field
(especially interesting are the $N=128$ and $N=192$
matrices obtained in Ref.~\cite{NishimuraTsuchiya2019}).

The analysis in the sections below will start from the  test-1
master field \eqref{eq:Ahat0-test} and \eqref{eq:Ahat1-test-1},
but, later, will also consider a test-2 master field
$\underline{\widehat{A}}^{\;1}_\text{\,test-2}$ with more structure.
Roughly speaking, this test-2 master field
again has a band-diagonal structure with parabolic behavior,
but now there is also
a finer modulation of $(2\,\Delta N)\times (2\,\Delta N)$ diagonal blocks,
which alternatingly are reduced by a positive factor $\kappa<1$
or boosted by a positive factor $\lambda>1$,
and a further modulation of $\Delta N\times \Delta N$ diagonal blocks,
which alternatingly have $+1$ or $-1$ on the diagonal.

Assuming $N$, $\Delta N$,  and $N/\Delta N \equiv L$
all to be even integers, this $N\times N$ traceless Hermitian
test matrix is given by the following expression:%
\bsubeqs\label{eq:Ahat1-test-2}
\beqa
\label{eq:Ahat1-test-2-Hermiteanized}
\underline{\widehat{A}}^{\;1}_\text{\,test-2}
&=&
\frac{1}{2}\;
\left(\widetilde{A}^{\;1}+\widetilde{A}^{\,1\;\dagger}\right)
-\frac{1}{N}\;\text{tr}\,
\left[\frac{1}{2}\;\left(\widetilde{A}^{\;1}
      +\widetilde{A}^{\,1\;\dagger}\right)\right]\,
\id_{\,N}
\,,
\eeqa
%%\\[2mm]
\beqa
\label{eq:Ahat1-test-2-diagonal-structure}
\widetilde{A}^{\;1}
&=&
D_{\kappa\lambda}
\cdot
D_\text{pm}
\cdot \overline{A}^{\;1}\,,
\eeqa
%%\\[2mm]
\beqa
\label{eq:Ahat1-test-2-diagonal-Dkappalambda}
D_{\kappa\lambda}
&=&
\text{diag}\big[
\kappa, \,\ldots\, ,\,  \kappa,\,
\lambda,\,\ldots\, ,\,  \lambda,\,
\,\ldots\,,\,
\lambda,\,\ldots\, ,\,  \lambda
\big]\,,
\eeqa
%%\\[2mm]
\beqa
\label{eq:Ahat1-test-2-diagonal-Dpm}
D_\text{pm}
&=&
\text{diag}\big[
+1, \,\ldots\, ,\,  +1,\,
-1,\,\ldots\, ,\,  -1,\,
\,\ldots\,,\,
-1, \,\ldots\, ,\,  -1
\big]\,,
\eeqa
%%\\[2mm]
\beqa
\label{eq:Ahat1-test-2-randomized-band}
\left(\overline{A}^{\;1} \right)_{i,\,j}
&=&
\begin{cases}
 \overline{r}_{\xi}\,\overline{x}^{\;1}(i) \,,   &
 \text{for}\;  j-\Delta N/2 \leq i \leq j+\Delta N/2\,,
 \\[2mm]
 0 \,,   &  \text{otherwise} \,,
\end{cases}\,,
\eeqa
%%\\[2mm]
\beqa
\label{eq:Ahat1-test-2-range}
\overline{r}_{\xi} &=& \text{rand}\big[1-\xi,\,1+\xi\big]
\,,
\eeqa
%%\\[2mm]
\beqa
\label{eq:Ahat1-test-2-x1bar}
\overline{x}^{\;1}(i)&=&
1 + \left[\left(i-\frac{1}{2}\right)\frac{2}{N} -1 \right]^2\,,
\eeqa
%%\\[2mm]
\beqa
\label{eq:Ahat1-test-2-kappa-lambda}
0 &<&  \kappa \leq 1 \leq \lambda\,,
\eeqa
%%\\[2mm]
\beqa
\label{eq:Ahat1-test-2-xi}
0 &<&  \xi < 1 \,,
\eeqa
\esubeqs
with $i,\,j \in \{1,\,\ldots ,\,  N\}$ and $\text{rand}\big[x,\,y\big]$
defined below \eqref{eq:Ahat0-test-alphatilde}.
The real numbers $\kappa$ and  $\lambda$
in \eqref{eq:Ahat1-test-2-diagonal-Dkappalambda}
are each repeated $2\,\Delta N$ times
[making for $L/2$ diagonal $(2\,\Delta N) \times (2\,\Delta N)$ blocks]
and the real numbers $+1$ and  $-1$
in \eqref{eq:Ahat1-test-2-diagonal-Dpm}
are each repeated $\Delta N$ times
[making for $L$ diagonal $\Delta N \times \Delta N$ blocks].

The $\pm 1$ fine-structure of \eqref{eq:Ahat1-test-2-diagonal-Dpm}
is inspired by the similar fine-structure of an
exact ``classical'' solution with \mbox{$\Delta N \sim 1$}
found in App.~A of Ref.~\cite{Klinkhamer2019-emergence-PTEP}.
The \textit{raison d'$\,\hat{\text{e}}$tre}
of the $\kappa,\,\lambda$ fine-structure
in \eqref{eq:Ahat1-test-2-diagonal-Dkappalambda} will become clear
in Sec.~\ref{sec:Spacetime-points-from-test-2-master-field}.
Remark also that the IIB-matrix-model variables are complex
Hermitian, whereas the two test master fields of this
section are real.

%%\newpage%%tmp
\section{Extraction procedure}
\label{sec:Extraction-procedure}

The procedure for obtaining spacetime points from the
master field has been outlined in Sec.~IV of Ref.~\cite{Klinkhamer2020}.
The basic idea is to consider,
in each of the ten matrices $\underline{\widehat{A}}^{\,\mu}$,
the $K$ blocks of size $n\times n$ centered on the diagonal.
Here, we assume that $N=K*n$, for positive integers $K$ and $n$,
and that $\underline{\widehat{A}}^{\,0}$ has already been
diagonalized and ordered by an appropriate global gauge
transformation.
The coordinates of the spacetime points are then obtained
from the average of the eigenvalues in each $n\times n$ block.
If the IIB-matrix-model master field $\underline{\widehat{A}}^{\,\mu}$
has a diagonal band width $\Delta N$,  we expect that $n$
must be chosen to be approximately equal to $\Delta N$ or,
better, significantly larger than $\Delta N$.
Furthermore, it is possible to introduce
a length scale $\ell$ into the IIB matrix model,
as discussed in Ref.~\cite{Klinkhamer2020},
so that the bosonic matrix variable $A^{\mu}$ carries the
dimension of length, as does the corresponding master field
$\underline{\widehat{A}}^{\,\mu}$.
Throughout this article, we take length units which set $\ell=1$.

The test master field $\underline{\widehat{A}}^{\;0}_\text{\,test}$,
as given by \eqref{eq:Ahat0-test}, then gives
the following temporal coordinate:%
\beq\label{eq:xhat0sigma}
\widehat{x}^{\;0}(\sigma)
\equiv
\widetilde{c}\;\widehat{t}(\sigma)
=
\frac{1}{n}\;\sum_{l=1}^{n} \, \overline{\alpha}_{(k-1)\,n+l}\,,
\eeq
for $\sigma\equiv k/K \in (0,\,1]$ with
$k \in \{1,\,\ldots ,\,  K\}$ and $K=N/n$.
The velocity $\widetilde{c}$ in \eqref{eq:xhat0sigma}
will be set to unity in the following.
The temporal coordinate $\widehat{t}^{\;1}(\sigma)$
takes values in the range $[-0.5,\,0.5]$.

Similarly,
the test-1 master field $\underline{\widehat{A}}^{\;1}_\text{\,test-1}$,
as given by \eqref{eq:Ahat1-test-1}, gives
the following coordinate in one spatial dimension:
\beq\label{eq:xhat1sigma}
\widehat{x}^{\;1}(\sigma)
=
\frac{1}{n}\;\sum_{l=1}^{n} \,
\left[\,\overline{\beta}^{\;1}\right]_{(k-1)\,n+l}\,,
\eeq
for $\sigma\equiv k/K \in (0,\,1]$
and eigenvalues $\overline{\beta}^{\;1}_{i}$
of the $n\times n$ blocks along the diagonal
in the $N\times N$  matrix  \eqref{eq:Ahat1-test-1}.
The same expression \eqref{eq:xhat1sigma} gives a spatial
coordinate from the test-2 master field \eqref{eq:Ahat1-test-2}.

Before we start with the explicit calculation
of the spacetime points \eqref{eq:xhat0sigma}
and \eqref{eq:xhat1sigma} from our test master fields,
we have a few general remarks on the adopted coarse-graining
procedure.
In order to cover the $\Delta N$ band diagonal
of the master field completely, it would perhaps be better
to let the $n\times n$ blocks, with $n > \Delta N$,
overlap halfway. But, then, there would be the problem
of double-counting some of the information
contained in the master field and, moreover, there would be
no way to obtain a clear ordering of the extracted information
with respect to time.
For these reasons, we prefer to keep, in our exploratory analysis,
the simple procedure of having the $n\times n$ blocks
touch on the diagonal, without overlap:
even though the $n\times n$ block eigenvalues are not perfect
(some information from the master field has been lost),
there is a clear ordering of the average \eqref{eq:xhat1sigma}
with respect to the coordinate time $t$ from \eqref{eq:xhat0sigma}.
%
%frk: keep tmp
%
%For fixed $N$ and $\Delta N$,
%the missed information from the master field would be reduced
%by taking a large value for the size $n$ the averaging blocks,
%$n \gg \Delta N$, but this would go at the price of reducing
%the number of spacetime points ($K=N/n$).

%%\newpage%%tmp
\section{Spacetime points from the test-1 master field}
\label{sec:Spacetime-points-from-test-1-master-field}

\begin{figure}[p]
\vspace*{-10mm}
\begin{center}
\includegraphics[width=0.95\textwidth]
{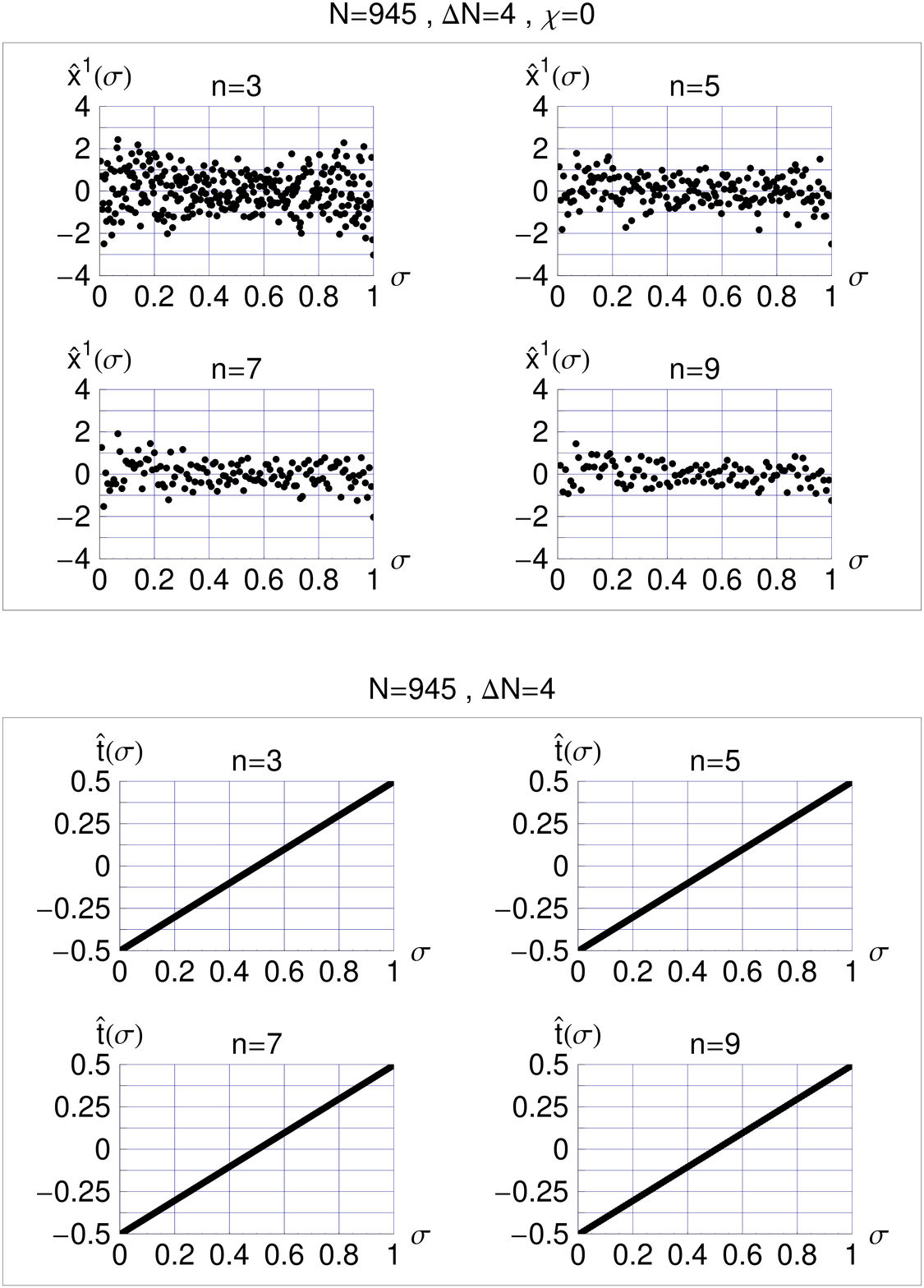}
%{xhat1-from-test-master-field-fig1-v090.eps}
%{xhat1-from-test-master-field-fig1-v040.eps}
%{xhat1-from-test-master-field-fig05-v020.eps}
\end{center}
\vspace*{-9mm}
\caption{The spatial coordinate $\widehat{x}^{\;1}(\sigma)$,
for $\sigma \in (0,\,1]$, is shown in the upper panel-quartet.
This coordinate $\widehat{x}^{\;1}(\sigma)$ is obtained by the procedure
of Sec.~\ref{sec:Extraction-procedure}
applied to the test-1 master field $\widehat{A}^{\;1}_\text{\,test-1}$
as given by \eqref{eq:Ahat1-test-1},
for matrix size $N=3*5*7*9=945$,
band-diagonal width $\Delta N=4$, and parameter $\chi=0$ to
select a continuous  range of values on the individual rows
of the band diagonal of the matrix.
The temporal coordinate $\widehat{t}^{\;1}(\sigma)$,
obtained by applying the same procedure to the 
matrix  \eqref{eq:Ahat0-test}, is shown in the lower panel-quartet.
Eliminating $\sigma$ between $\widehat{x}^{\;1}(\sigma)$
and $\widehat{t}(\sigma)$ gives $\widehat{x}^{\;1}\big(\,\widehat{t}\:\big)$,
which is shown in Fig.~\ref{fig:fig2}.
}
\label{fig:fig1}%%{fig:fig05}
%\vspace*{100mm}
\end{figure}

Now, choose fixed values of $N$ (assumed to be odd)
and $\Delta N$ (assumed to be even) in
the test master field $\widehat{A}^{\;1}_\text{\,test-1}$
from \eqref{eq:Ahat1-test-1} for $\chi=0$.
Then, for various choices of the block size $n$
(which must be odd, because $N$ has been assumed to be odd),
the procedure from Sec.~\ref{sec:Extraction-procedure}
gives the spatial coordinate $\widehat{x}^{\;1}(\sigma)$
from \eqref{eq:xhat1sigma}, for $\sigma \in (0,\,1]$.
Numerical results are presented
in the upper panel-quartet of Fig.~\ref{fig:fig1}.
All numerical results reported in this paper were obtained
with \textsc{Mathematica} 5.0~\cite{Wolfram1991}.
See, in particular, Sec.~3.2.3 of Ref.~\cite{Wolfram1991} 
for the use of pseudorandom numbers in \textsc{Mathematica} and
Chap. 3 of Ref.~\cite{Knuth1998} for a general discussion.

\begin{figure}[p]
\vspace*{-5mm}
\begin{center}
\includegraphics[width=0.95\textwidth]
{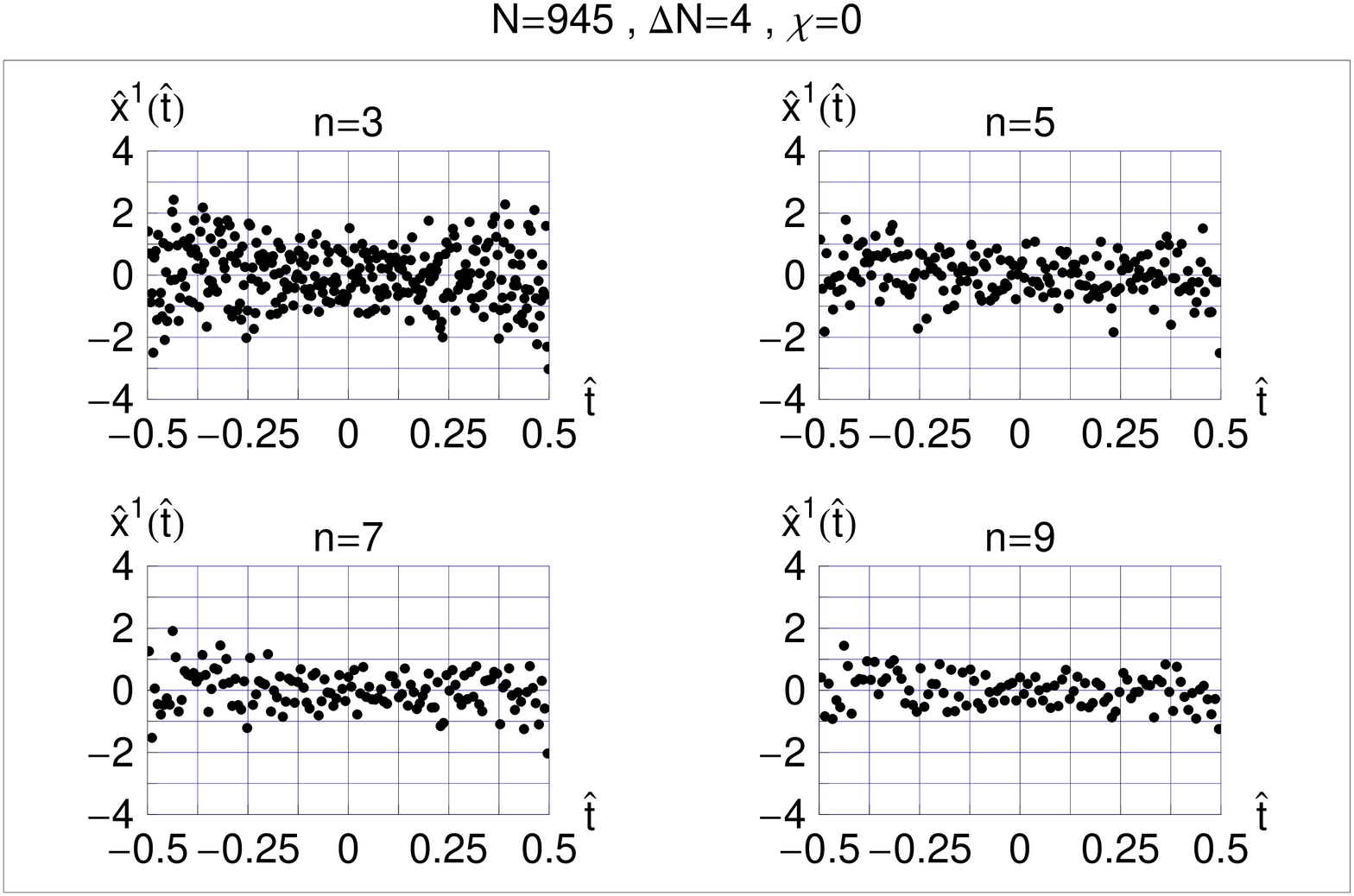}
%{xhat1-from-test-master-field-fig2-v090.eps}
%{xhat1-from-test-master-field-fig2-v040.eps}
%{xhat1-from-test-master-field-fig06-v020.eps}
\end{center}
\vspace*{-5mm}
\caption{Behavior of $\widehat{x}^{\;1}$ versus $\widehat{t}$
from the test-1 results of Fig.~\ref{fig:fig1}.}
\label{fig:fig2}%%{fig:fig06}
%\vspace*{100mm}
%\end{figure}
\vspace*{5mm}
%\begin{figure}[p]
\vspace*{-0mm}
\begin{center}
\includegraphics[width=0.95\textwidth]
{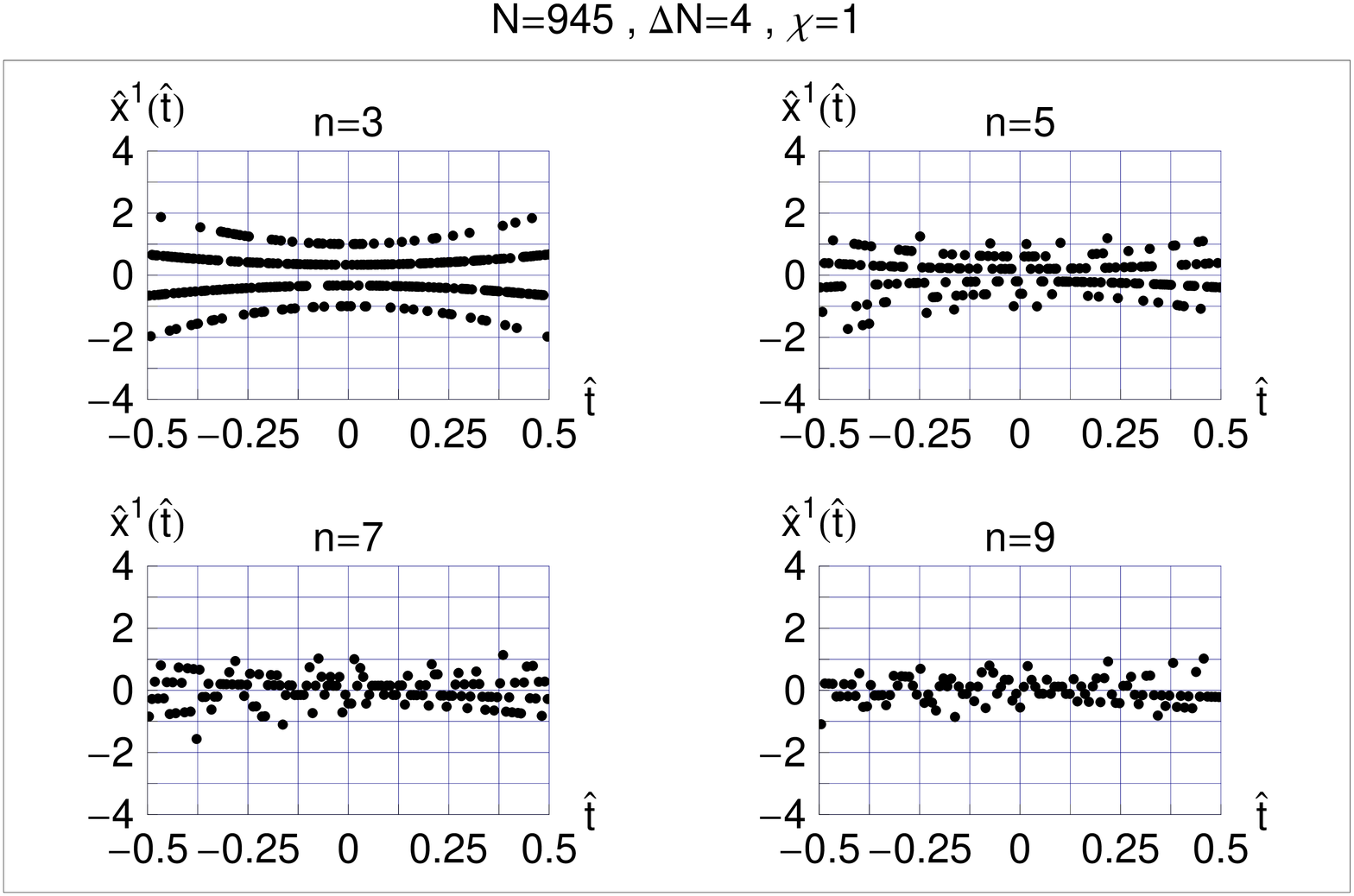}
%{xhat1-from-test-master-field-fig3-v090.eps}
%{xhat1-from-test-master-field-fig3-v040.eps}
%{xhat1-from-test-master-field-fig07-v020.eps}
\end{center}
\vspace*{-5mm}
\caption{Same as Fig.~\ref{fig:fig2}, but the test-1 master field
\eqref{eq:Ahat1-test-1} now has parameter $\chi=1$ to select
a discrete range of values on the individual rows
of the band diagonal of the matrix.
}
\label{fig:fig3}%%{fig:fig07}
%\vspace*{100mm}
\end{figure}

The calculation of the temporal coordinate is simpler, as
the test master field $\underline{\widehat{A}}^{\;0}_\text{\,test}$
from \eqref{eq:Ahat0-test} is already diagonal
with explicit eigenvalues.
The temporal coordinate $\widehat{t}(\sigma)$ follows from
\eqref{eq:xhat0sigma} for $\widetilde{c}=1$ and $\sigma\in (0,\,1]$.
It turns out that $\widehat{t}$
is approximately linearly proportional to $\sigma$,
as shown by the lower panel-quartet of Fig.~\ref{fig:fig1}.
From these results we obtain $\widehat{x}^{\;1}\big(\,\widehat{t}\:\big)$,
as shown by Fig.~\ref{fig:fig2}.

For completeness, we also show, in Fig.~\ref{fig:fig3},
the results with discrete
values ($\chi=1$) on the rows of the band diagonal for $\Delta N=4$,
where the role of the $\chi$ parameter has been explained in the
sentence starting below \eqref{eq:Ahat1-test-1-chi}.
Similar $n=3$ results have been obtained for $\Delta N=2$ and $\Delta N=6$.

\begin{figure}[p]
\vspace*{-5mm}
\begin{center}
\includegraphics[width=0.95\textwidth]
{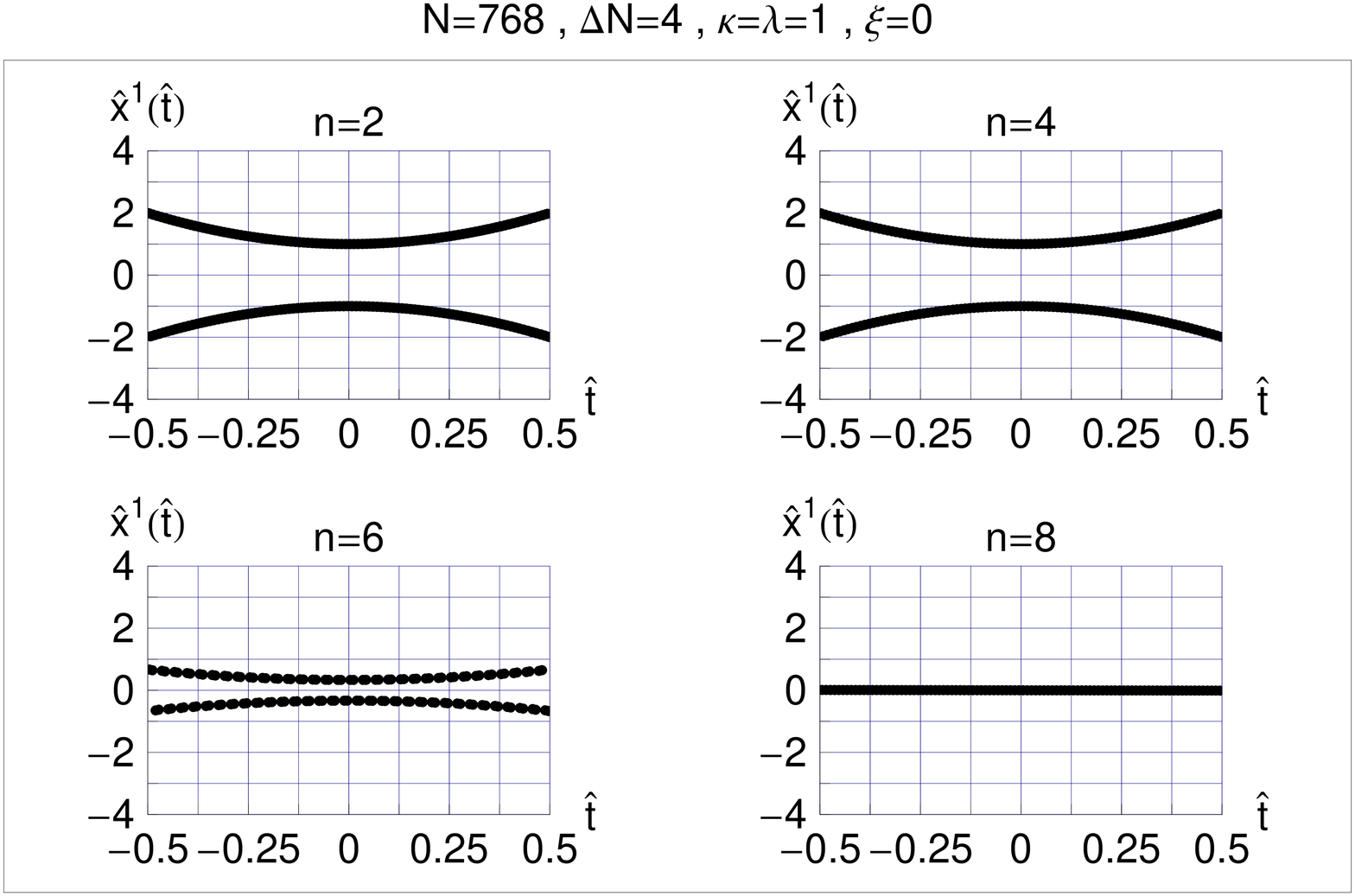}
%{xhat1-from-test-master-field-fig4-v050.eps}
\end{center}
\vspace*{-5mm}
\caption{Behavior of $\widehat{x}^{\;1}$ versus $\widehat{t}\,$ from
the test-2 master field \eqref{eq:Ahat1-test-2},
for $N=2^{8}*3=768$ and $\Delta N=4$, and with trivial modulation parameters,
$\kappa=\lambda=1$, and vanishing randomization parameter, $\xi=0$.}
\label{fig:fig4}
\vspace*{100mm}
\end{figure}

\begin{figure}[p]
\vspace*{-5mm}
\begin{center}
\includegraphics[width=0.95\textwidth]
{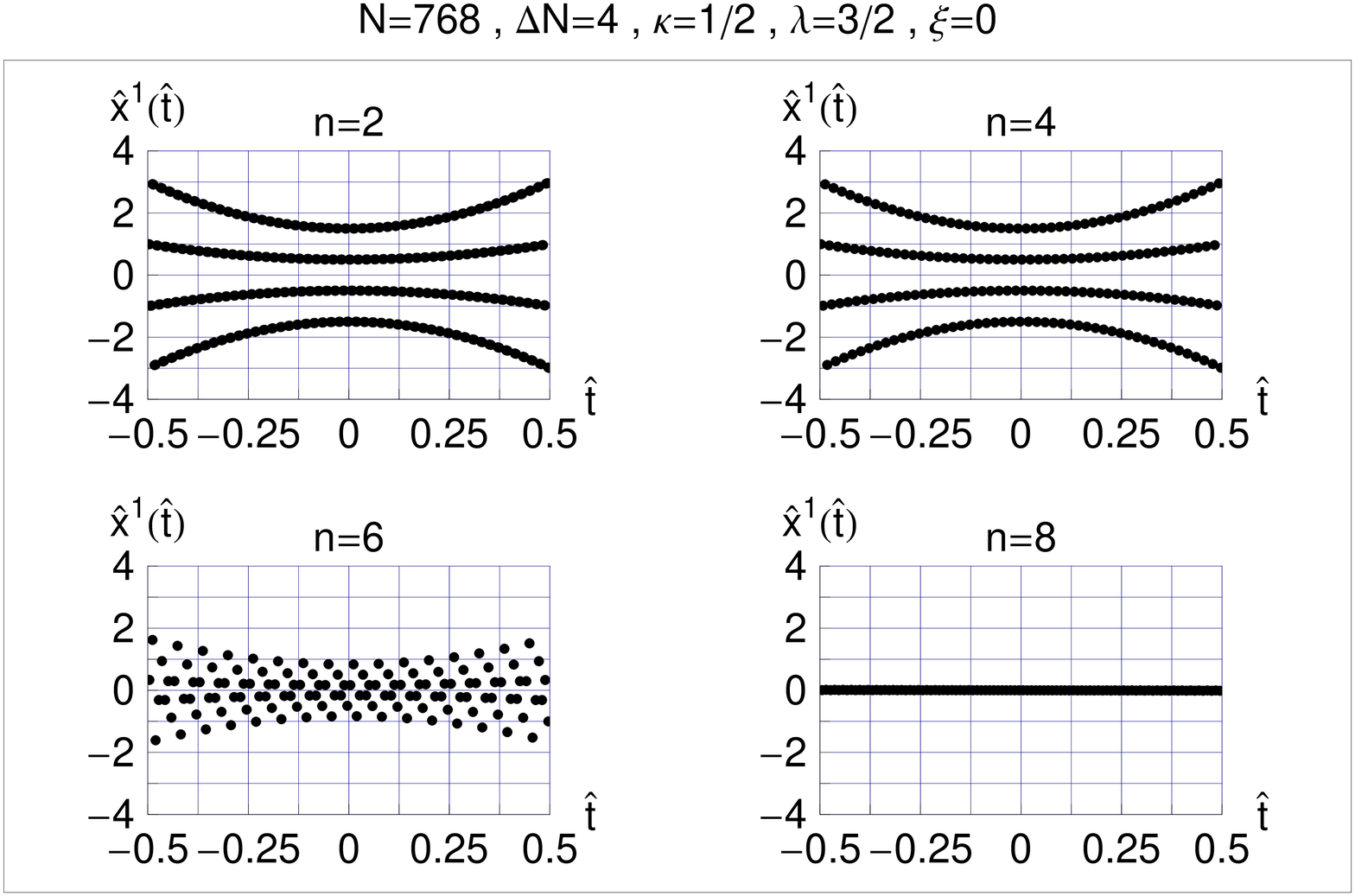}
%{xhat1-from-test-master-field-fig5-v050.eps}
\end{center}
\vspace*{-5mm}
\caption{Same as Fig.~\ref{fig:fig4}, but now with
nontrivial modulation parameters, $\kappa=1/2$ and $\lambda=3/2$.}
\label{fig:fig5}
%\vspace*{0mm}
%\end{figure}
\vspace*{10mm}
%\begin{figure}[p]
\vspace*{-0mm}
\begin{center}
\includegraphics[width=0.95\textwidth]
{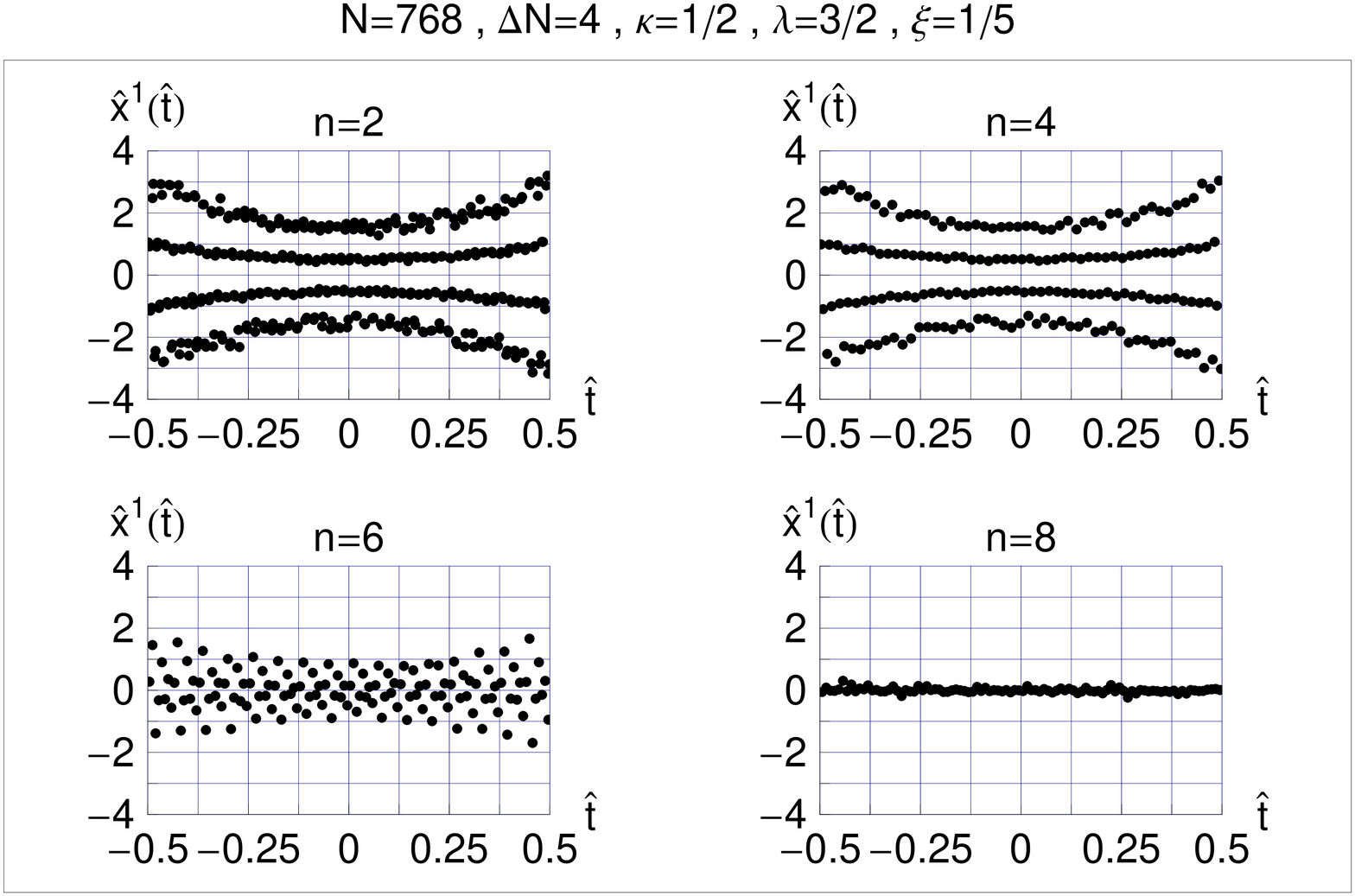}
%{xhat1-from-test-master-field-fig6-v050.eps}
\end{center}
\vspace*{-5mm}
\caption{Same as Fig.~\ref{fig:fig5}, but now with a nonvanishing
randomization parameter, $\xi=0.2$.}
\label{fig:fig6}
%\vspace*{100mm}
\end{figure}

\newpage
\section{Spacetime points from the test-2 master field}
\label{sec:Spacetime-points-from-test-2-master-field}

The $n=3$ results of Fig.~\ref{fig:fig2} perhaps show
a point set expanding with time $|t|$, but it is not clear
if a classical spacetime emerges with an  expanding \emph{volume}
(at this moment, we do not
have the metric distance between the points).
In that sense, the $n=3$ results of Fig.~\ref{fig:fig3}
may be more promising, as they \emph{suggest}
four ``strands'' of spacetime separating from each other
as $|t|$ increases
(we are using ``strand'' in the meaning of a ``strand of pearls'').
Incidentally, a somewhat related pattern
has been observed
in Fig.~5 of Ref.~\cite{Hatakeyama-etal2020}.

Returning to the strands of Fig.~\ref{fig:fig3},
the following question arises:
is it at all possible to \emph{design} a special master field
$\underline{\widehat{A}}^{\;1}_\text{\,special}$,
so that the procedure of Sec.~\ref{sec:Extraction-procedure}
gives \emph{multiple} strands of spacetime,
which fill out an expanding universe?

The answer is affirmative and an example
for up to four spacetime strands is given by
the test-2 matrix \eqref{eq:Ahat1-test-2}.
For $\kappa=\lambda=1$, we get two strands
(Fig.~\ref{fig:fig4})
and, for $\kappa=1/2$ and $\lambda=3/2$,
for example, we get four strands with $n=2$ and $n=4$ averaging
(Figs.~\ref{fig:fig5} and \ref{fig:fig6});
the $n=6$ results of Fig.~\ref{fig:fig5} even suggest
the presence of six strands.
It appears possible to get more than four (or six) spacetime strands
by introducing even more parameters and structure along the diagonal
of the test matrix.

The main open problem, now, is to obtain the emergent metric
and to calculate the metric
distances between points on a single strand and
between points on different strands.
At this moment, we have no solid results but only
some speculative remarks.
One speculative remark is that, for two strands
extending in two ``large'' spatial dimensions,
two ``neighboring'' points on a single strand may have a
smaller metric distance than two ``neighboring'' points
on different adjacent strands.
Appendix~\ref{app:Toy-model-calculation-metric-distances}
reports on a toy-model calculation which suggests such a result.

%%\newpage%%tmp
\section{Discussion}
\label{sec:Discussion}

In this somewhat technical paper, we have considered several test
matrices and obtained tentative spacetime points by
applying the procedure of Ref.~\cite{Klinkhamer2020} to these matrices
(an alternative procedure is presented in
App.~\ref{app:Fuzzy-sphere-and-alternative-procedure}). 
These test matrices have a band-diagonal structure, one
being strictly diagonal to represent the time coordinate $t$
and another having a finite bandwidth $\Delta N$
to represent a typical spatial coordinate $x^{1}$
from a ``large'' dimension,
whose average absolute value $|x^{1}|$ grows quadratically with $t$
(a behavior seen in the numerical results of
Refs.~\cite{KimNishimuraTsuchiya2012,NishimuraTsuchiya2019,Hatakeyama-etal2020}).
The hope is that these test matrices may help us to understand
a possible fine-structure of the genuine IIB-matrix-model
master field.

As a first step towards such an understanding, we have
constructed the test-2 matrix from \eqref{eq:Ahat1-test-2},
which has a very special fine-structure to allow for the appearance
of multiple ``strands'' of spacetime
(Figs.~\ref{fig:fig4}--\ref{fig:fig6}).
In fact, this understanding allows us to interpret the
somewhat surprising $n=3$ and $n=5$ results of Fig.~\ref{fig:fig3},
which indicate the appearance of, respectively, four and six strands
(with some good will, the $n=7$ results can be seen to
hint at the presence of eight strands).
The heuristic idea, now, is that the simple flip-flop behavior
on each row of the matrix gives a repeating pattern if a sufficiently
large number of rows is considered ($N\to\infty$).
Preliminary numerical results extending the calculation of
Fig.~\ref{fig:fig3} appear to confirm the appearance of more than
six strands.

Many questions remain as to the procedure
for the extraction of the spacetime points,
not to mention the spacetime metric.
But even more important, at this moment, is to obtain
a \emph{reliable} approximation of the IIB-matrix-model master field,
which may or may not display some form of fine-structure.

%%%%%%\newpage%%tmp
%\vspace*{-5mm}
\begin{acknowledgments}  
%\vspace*{-5mm}
It is a pleasure to thank H. Steinacker
for helpful discussion on the fuzzy sphere.
%\vspace*{-0mm}
\end{acknowledgments}

%%\newpage%%tmp
\begin{appendix}
\section{Toy-model calculation of metric distances}
\label{app:Toy-model-calculation-metric-distances}

In this appendix, we report on a toy-model calculation
which extends a similar calculation presented in App.~B %Sec.~V
of Ref.~\cite{Klinkhamer2020}.

Start from the emergent inverse metric as given
by (5.1) of Ref.~\cite{Klinkhamer2020}
(based on an earlier expression from  Ref.~\cite{Aoki-etal-review-1999}):
\beq \label{eq:emergent-inverse-metric-app}
g^{\mu\nu}(x) \sim
\int_{\mathbb{R}^{10}} d^{10}y\;
\langle\langle\, \rho(y)  \,\rangle\rangle
\; (x-y)^{\mu}\,(x-y)^{\nu}\;f(x-y)\;r(x,\,y)\,,
\eeq
where $\langle\langle\, \rho(x)  \,\rangle\rangle$
is the average density function of emergent spacetime
points,
$r(x,\,y)$ is the correlation function from these density functions,
and $f(x-y)$ is a correlation function that appears
in the effective action of a low-energy scalar
degree of freedom; see Refs.~\cite{Aoki-etal-review-1999,%
Klinkhamer2020} for further details.

\begin{figure}[t]
\vspace*{-9mm}
\begin{center}
\includegraphics[width=0.5\textwidth] %%fig7-v3.eps=fig7-v4.eps
{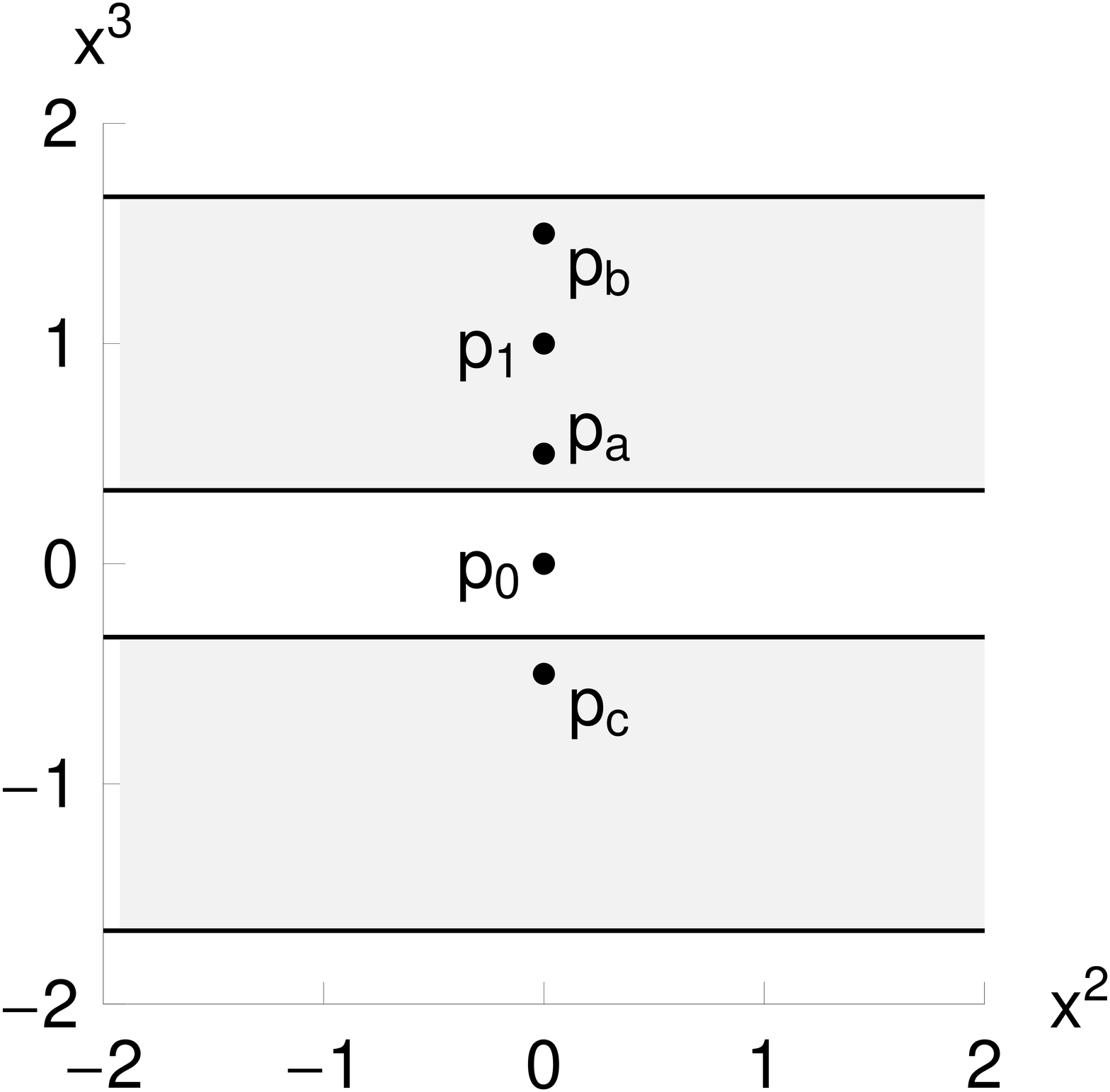}
%%{IIB-matrix-model-Extracting-spacetime-points-fig7-v3.eps}
%%{IIB-matrix-model-Extracting-spacetime-points-fig7-v202.eps}
%%{IIB-matrix-model-Extracting-spacetime-points-fig7-v2.eps}
%%{IIB-matrix-model-Extracting-spacetime-points-fig7-v180-60pt.eps}
\end{center}
\vspace*{-7mm}
\caption{Setup for the toy-model calculation, where the shaded
areas indicate the bands $B_{\pm}$ (centered on $x^3=\pm 1$ and
having widths $\Delta x = 4/3$) over which spacetime points emerge.
Metric components at two points $p_{0}$ and $p_{1}$ are compared
and metric distances between three further
points $p_{a}$, $p_{b}$, and $p_{c}$ calculated.
}
\label{fig:fig7}
\vspace*{0mm}
\end{figure}

%%XXX
Restrict  \eqref{eq:emergent-inverse-metric-app} to two ``large''
spatial dimensions (with coordinates $x^2$ and $x^3$) and
consider a simple setup in the $(x^2,\,x^3)$ plane with
two finite bands $B_{+}$ and $B_{-}$ (Fig.~\ref{fig:fig7}),
\bsubeqs\label{eq:B1-B2-app}
\beqa
B_{+} &=& \big\{ (x^2,\,x^3)\,\big|\,
x^2 \in [-2,\,2]\;,\;x^3 \in [+1-\Delta x/2,\,+1+\Delta x/2]\big\}\,,
\\[2mm]
B_{-} &=& \big\{ (x^2,\,x^3)\,\big|\,
x^2 \in [-2,\,2]\;,\;x^3 \in [-1-\Delta x/2,\,-1+\Delta x/2]\big\}\,,
\\[2mm]
0 &<& \Delta x/2 < 1\,,
\eeqa
\esubeqs
where the bands may be thought to arise from wiggly strands such as
shown by the $n=4$ results of Fig.~\ref{fig:fig6}.
The average density of emerging spacetime points
is assumed to be nonvanishing only over these bands,
\beq\label{eq:rho-Bplus-B2minus-app}
\langle\langle\, \rho(x)  \,\rangle\rangle
=
\begin{cases}
 1/\ell^{10} \,,   &
 \text{for}\;  (x^2,\,x^3) \in B_{+} \cup  B_{-}\,,
 \\[2mm]
 0 \,,   &  \text{otherwise} \,,
\end{cases}\,,
\eeq
with $\ell$ the length scale of the IIB matrix model
and all coordinates $x^{\mu}$ carrying the dimension of length,
as discussed in Sec.~\ref{sec:Extraction-procedure}.
The length units of the setup in Fig.~\ref{fig:fig7}
are assumed to correspond to $\ell = 1$.

Next, introduce a symmetric cutoff on the integrals at $\pm 2$
and make two further simplifications,
\bsubeqs\label{eq:f-r-assumptions-app}
\beqa
\label{eq:f-assumptions-app}
f(x) &=& 1/\ell^2\,,
\\[2mm]
\label{eq:r-assumptions-app}
r(x,\,y) &=& 1\,,
\eeqa
\esubeqs
with $\ell=1$ for the chosen length units.
Note that the assumptions \eqref{eq:f-r-assumptions-app}
perhaps make sense if the space patch
considered is sufficiently small in physical units;
see also the discussion below for another setup.

It is, now, straightforward to calculate the inverse metric
components $g^{22}$, $g^{33}$, and $g^{23}=g^{32}$
from \eqref{eq:emergent-inverse-metric-app} for two
points $p_{0}$ and $p_{1}$ (Fig.~\ref{fig:fig7}) with the following
coordinates:
\bsubeqs\label{eq:p0-p1-app}
\beqa
(x^2,\,x^3)_{p_{0}} &=&  (0,\,0)\,,
\\[2mm]
(x^2,\,x^3)_{p_{1}} &=&  (0,\,1)\,,
\eeqa
\esubeqs
so that point $p_{1}$ lies in the upper band $B_{+}$
and point $p_{0}$ between the bands $B_{+}$ and $B_{-}$.
As the $y^3$ integral in \eqref{eq:emergent-inverse-metric-app}
is elementary, we obtain immediately
\bsubeqs\label{eq:inverse-metric-inequalities-app}
\beqa
0 &<& g^{33}(p_0) < g^{33}(p_1)\,,
\\[2mm]
0 &<& g^{22}(p_0) = g^{22}(p_1) \,,
\\[2mm]
0 &=& g^{23}(p_0) = g^{23}(p_1)\,.
\eeqa
\esubeqs
We actually need the metric
$g_{\mu\nu}(x)$ for calculating the metric distances
(from $ds^2 = g_{\mu\nu}\,dx^{\mu}\, dx^{\nu}$)
and we have the following inequalities and equalities
from \eqref{eq:inverse-metric-inequalities-app}:%
\bsubeqs\label{eq:metric-inequalities-p0-p1-app}
\beqa
\label{eq:metric-33-inequalities-p0-p1-app}
g_{33}(p_0) &>& g_{33}(p_1) >0\,,
\\[2mm]
g_{22}(p_0) &=& g_{22}(p_1)  >0\,,
\\[2mm]
g_{23}(p_0) &=& g_{23}(p_1) = 0\,.
\eeqa
\esubeqs
The results \eqref{eq:inverse-metric-inequalities-app}
and \eqref{eq:metric-inequalities-p0-p1-app}
extend to other points $\overline{p}_1$
with $x^2=0$ lying in the band $B_{+}$
and other points $\overline{p}_0$ with $x^2=0$
lying between the bands $B_{+}$ and $B_{-}$.
In fact, the metric component $g_{33}(x^2,\,x^3)$ at $x^2=0$
is a positive symmetric function of $x^3$ having
a maximum at $x^3=0$ and dropping monotonically with $|x^3|$,
\beq\label{eq:metric-33-inequalities-app}
g_{33}(0,\,|x^3|) > g_{33}(0,\,|x^{3\,\prime}|) > 0\,,
\;\;\;\text{for }\;\;\;0< |x^3| <|x^{3\,\prime}|\,.
\eeq

The results \eqref{eq:metric-33-inequalities-p0-p1-app}
and \eqref{eq:metric-33-inequalities-app}
are definitely surprising, with a larger metric component $g_{33}$
in the ``empty space'' between the bands than in the bands themselves.
These results trace back to the quadratic behavior of the $\mu=\nu=3$
integrand in \eqref{eq:emergent-inverse-metric-app},
as there is no damping from the remaining factors $f(x-y)$ and $r(x,\,y)$,
according to assumptions \eqref{eq:f-r-assumptions-app}.

A different setup has \eqref{eq:f-assumptions-app}
replaced by, for example,
\beq\label{eq:f-other setup-app}
f(x)=\ell^2\big/\big[\ell^4+(x^2)^2\big]\,,
\eeq
with $x^2 \equiv \eta_{\mu\nu}\,x^{\mu}\,x^{\nu}$,
in terms of the ``coupling constants'' $\eta_{\mu\nu}$
from the Lorentzian IIB matrix model.
In length units with $\ell=1$, the \textit{Ansatz}
\eqref{eq:f-other setup-app} simply reads
$f(x)=1/(1+x^4)$. This alternative setup
gives the inequality $g_{33}(p_0) < g_{33}(p_1)$,
which corresponds to having
a smaller metric component $g_{33}$ in the ``empty space''
between the bands than in the bands themselves.
Moreover, the actual value of $g_{33}(p_0)$
obtained from this setup with $f(x)=1/(1+x^4)$
is generically larger than the value obtained
from the original setup with $f(x)=1$.
This larger value of $g_{33}$  is
simply due to the reduction of the $\mu=\nu=3$ integrand
\eqref{eq:emergent-inverse-metric-app} by the
function $f(x)=1/(1+x^4)$, making for a smaller value of
$g^{33}$ and, hence, a larger value of $g_{33}$
(the same argument applies to the $g_{22}$ component).
A straightforward calculation of the squared distance
between the points $p_0$ and $p_1$ gives, in fact, a
larger value from the setup with $f(x)=1/(1+x^4)$
than from the setup with $f(x)=1$, all other assumptions being equal.
Together with the above parenthetical remark
on the $g_{22}$ component, we then find that
the space patch from the setup with $f(x)=1/(1+x^4)$
is larger than the space patch from the setup with $f(x)=1$.

Returning to the original setup with
assumptions \eqref{eq:f-r-assumptions-app}, we like to
give an explicit example and take the following
numerical value of the bandwidth:
\beq\label{eq:Delta-x-num-app}
\Delta x = 4/3\,,
\eeq
and consider the following three points $p_{a}$, $p_{b}$, and $p_{c}$
(Fig.~\ref{fig:fig7}) with coordinates:
\bsubeqs\label{eq:pa-pb-pc-app}
\beqa
(x^2,\,x^3)_{p_{a}} &=&  (0,\,1/2)\,,
\\[2mm]
(x^2,\,x^3)_{p_{b}} &=&  (0,\,3/2)\,,
\\[2mm]
(x^2,\,x^3)_{p_{c}} &=&  (0,\,-1/2)\,,
\eeqa
\esubeqs
so that $p_{a}$ and $p_{b}$ lie on the single band $B_{+}$
and $p_{c}$ on the other band $B_{-}$.
The coordinate distance between $p_{a}$ and $p_{b}$ equals
the coordinate distance between $p_{a}$ and $p_{c}$.
But, from \eqref{eq:metric-33-inequalities-app}, we obtain
different metric distances between these two pairs of points:
\beq\label{eq:dab-dac-app}
0 < d\left(p_{a},\,p_{b}\right) < d\left(p_{a},\,p_{c}\right)\,.
\eeq
If all assumptions hold true (possibly for
a relatively small region of spacetime), the
result \eqref{eq:dab-dac-app} would imply that
two spatially-separated neighboring points on a single band
have a smaller metric distance
than two spatially-separated neighboring points
on different adjacent bands.

%%\newpage%%tmp
\section{Fuzzy sphere and an alternative extraction procedure}
\label{app:Fuzzy-sphere-and-alternative-procedure}

\subsection{Fuzzy-sphere matrices}
\label{subapp:Fuzzy-sphere-matrices}

The so-called fuzzy sphere $S_{2}^{N}$ provides 
a relatively simple example of noncommutative geometry; see
two reviews~\cite{Hoppe2002,Steinacker2011}
for details and further references.
An explicit realization of the corresponding three matrices is given
by a particular irreducible representation of the Lie group $SU(2)$,
characterized by a fixed integer or odd-half-integer
quantum number $j$ and a variable quantum number $m$ 
ranging over $\{-j,\, -j+1,\,\ldots\,,\, j\}$:
\bsubeqs\label{eq:fuzzy-sphere-matrices-M-def-N}
\beqa
\left(M^a\right)^{(j)}_{m,m'}  
&=& \langle j,m'| X^a  |j,m \rangle\,,
\quad
\text{for}  
\quad
a \in \{1,\, 2,\,3\}\,,
\\[2mm]
N&=&2\,j+1 \,,
\eeqa
\esubeqs
with the following definitions:
\bsubeqs\label{eq:fuzzy-sphere-matrices-construction}
\beqa
X^{1}  &=&  \left(X_{+} + X_{-}\right)/2\,,
\\[2mm]
X^{2}  &=&  \left(X_{+} - X_{-}\right)/(2\,i)\,,
\\[2mm]
X_{\pm}  &=&  X^1   \pm i\,X^2\,,
\\[2mm]
\label{eq:fuzzy-sphere-matrices-construction-X3}
X^3 \,|j,m \rangle  &=&  m \,|j,m \rangle\,,
\\[2mm]
\label{eq:fuzzy-sphere-matrices-construction-Xplus}
X_{+} \,|j,m \rangle  &=&  r_{+} \,|j,m+1 \rangle\,,
\\[2mm]
\label{eq:fuzzy-sphere-matrices-construction-Xmin}
X_{-} \,|j,m \rangle  &=&  r_{-} \,|j,m-1 \rangle\,,
\\[2mm]
\label{eq:fuzzy-sphere-matrices-construction-rpm}
r_{\pm}  &\equiv& \sqrt{(j \mp m)\,(j \pm m+1)}\,.
\eeqa
\esubeqs
In this way, we obtain three $N\times N$ matrices $M^a$, 
which satisfy the $su(2)$ Lie algebra
\beq
\label{eq:algebra}
\left[M^a,\,M^b\right]  = i\,\epsilon_{abc}\,M^c\,,
\eeq
with Levi--Civita symbol $\epsilon_{abc}$ (normalized as
$\epsilon_{123}=1$), and which have the norm square
\bsubeqs\label{eq:sum-M-squares-def-CN}
\beqa\label{eq:sum-M-squares} 
M^1 \cdot M^1 + M^2 \cdot M^2 +M^3 \cdot M^3 &=& C_N\, \id_N\,,
\\[2mm]
\label{eq:def-CN}
C_N &\equiv& \frac{1}{4}\, \left(N^2-1\right)\,.
\eeqa
\esubeqs
For later discussion, we explicitly give the 
traceless Hermitian  matrices for $N=9\,$:
\bsubeqs\label{eq:M3-M1-M2-N9}
\beqa
M^3 &=&
\left(
\renewcommand{\arraycolsep}{0.5pc} %% enlarge column spacing 
\renewcommand{\arraystretch}{1.0} %% enlarge line spacing
\begin{array}{ccccccccc}
-4 & 0 & 0 & 0 & 0 & 0 & 0 & 0 & 0 \\ 
0 &  -3 & 0 & 0 & 0 & 0 & 0 & 0 & 0 \\ 
0 & 0 &  -2 & 0 & 0 & 0 & 0 & 0 & 0 \\ 
0 & 0 & 0 & -1  & 0 & 0 & 0 & 0 & 0 \\ 
0 & 0 & 0 & 0   & 0 & 0 & 0 & 0 & 0 \\ 
0 & 0 & 0 & 0   & 0 & 1 & 0 & 0 & 0 \\ 
0 & 0 & 0 & 0   & 0 & 0 & 2 & 0 & 0 \\ 
0 & 0 & 0 & 0   & 0 & 0 & 0 & 3 & 0 \\ 
0 & 0 & 0 & 0   & 0 & 0 & 0 & 0 & 4 \\ 
\end{array}
\right)\,,
\eeqa
%\\[4mm]
\beqa
M^1 &=&   
\left(
\renewcommand{\arraycolsep}{0.375pc} %% enlarge column spacing 
\renewcommand{\arraystretch}{1.0} %% enlarge line spacing
\begin{array}{ccccccccc}
0 & {\sqrt{2}} & 0 & 0 & 0 & 0 & 0 & 0 & 0 \\ 
{\sqrt{2}} & 0 &\sqrt{7/2} & 0 & 0 & 0 & 0 & 0 & 0 \\ 
0 &\sqrt{7/2} & 0 & 3/\sqrt{2} & 0 & 0 & 0 & 0 & 0 \\ 
0 & 0 & 3/\sqrt{2} & 0 & {\sqrt{5}} & 0 & 0 & 0 & 0 \\ 
0 & 0 & 0 & {\sqrt{5}} & 0 & {\sqrt{5}} & 0 & 0 & 0 \\ 
0 & 0 & 0 & 0 & {\sqrt{5}} & 0 & 3/\sqrt{2} & 0 & 0 \\ 
0 & 0 & 0 & 0 & 0 & 3/\sqrt{2} & 0 &\sqrt{7/2} & 0 \\ 
0 & 0 & 0 & 0 & 0 & 0 &\sqrt{7/2} & 0 & {\sqrt{2}} \\ 
0 & 0 & 0 & 0 & 0 & 0 & 0 & {\sqrt{2}} & 0 \\ 
\end{array}
%\end{tabular} 
\right)\,,
\eeqa
%\\[4mm]
\beqa
M^2 &=& i\,
\left(
  \begin{array}{ccccccccc}
0 & {\sqrt{2}} & 0 & 0 & 0 & 0 & 0 & 0 & 0 \\ 
-{\sqrt{2}} & 0 & \sqrt{7/2} & 0 & 0 & 0 & 0 & 0 & 0 \\ 
0 & -\sqrt{7/2} & 0 & 3/\sqrt{2} & 0 & 0 & 0 & 0 & 0 \\ 
0 & 0 & -3/\sqrt{2} & 0 & {\sqrt{5}} & 0 & 0 & 0 & 0 \\ 
0 & 0 & 0 & -{\sqrt{5}} & 0 & {\sqrt{5}} & 0 & 0 & 0 \\ 
0 & 0 & 0 & 0 & -{\sqrt{5}} & 0 & 3/\sqrt{2} & 0 & 0 \\ 
0 & 0 & 0 & 0 & 0 & -3/\sqrt{2} & 0 & 
\sqrt{7/2} & 0 \\ 
0 & 0 & 0 & 0 & 0 & 0 & -\sqrt{7/2} & 0 & {\sqrt{2}} \\ 
0 & 0 & 0 & 0 & 0 & 0 & 0 & -{\sqrt{2}} & 0 \\
  \end{array}
\right)\,.
\eeqa
\esubeqs

As it stands, the matrices $M^a$ from
\eqref{eq:fuzzy-sphere-matrices-M-def-N}--\eqref{eq:fuzzy-sphere-matrices-construction}
are dimensionless.
But they can be given the dimension of length if we add
a length scale $\ell$ on the right-hand sides of 
\eqref{eq:fuzzy-sphere-matrices-construction-X3}--%
\eqref{eq:fuzzy-sphere-matrices-construction-Xmin}
and \eqref{eq:M3-M1-M2-N9}. There is then a factor
$1/\ell$ on the right-hand side of \eqref{eq:algebra}
and a factor $\ell^2$ on the right-hand side of \eqref{eq:sum-M-squares}.
Taking appropriate length units to set $\ell=1$,
the above equations hold as given.

%%\newpage%%tmp
\subsection{Alternative extraction procedure}
\label{subapp:Alternative-extraction-procedure}

The matrices of the fuzzy-sphere or, more generally,
the matrices of noncommutative geometry may, at best, have an indirect 
relevance to the matrices of the IIB-matrix-model master field
(see, e.g., Refs.~\cite{Connes2019,Steinacker2019} for two recent
reviews on noncommutative geometry).
Here, we only intend to use the fuzzy-sphere matrices as one possible 
test bench for the extraction procedure of spacetime points.
Recall that the extraction procedure is ultimately to be applied to
the exact IIB-matrix-model master-field matrices. 

For the fuzzy-sphere matrices 
\eqref{eq:fuzzy-sphere-matrices-M-def-N}--\eqref{eq:fuzzy-sphere-matrices-construction}
with $a=1$ or $a=2$, we note that the extraction procedure of 
Sec.~\ref{sec:Extraction-procedure} is unsatisfactory.
The eigenvalues of the diagonal $n\times n$ blocks, 
for even $n\geq 2$ and $N$ an integer multiple of $n$,  
come in pairs of opposite values
and the eigenvalues of the  diagonal $n\times n$ blocks, 
for odd $n\geq 3$ and $N$ an integer multiple of $n$,  
come in pairs of opposite values or as eigenvalue zero.
This implies that the averages \eqref{eq:xhat1sigma} simply give zero
(see below for an explicit example with $N=9$ and $n=3$).

But there are alternative procedures. One procedure is to 
consider again the \emph{adjacent} (non-overlapping) $n\times n$ 
blocks along the diagonals of the gauge-transformed
master-field matrices  $\underline{\widehat{A}}^{\,\mu}$
(here, the matrices $M^a$),
but now to randomly take from each block a \emph{single}
eigenvalue, making for the discrete points $\widetilde{x}^{\,\mu}_{k}$
(here, the points $\widetilde{x}^{\,a}_{k}$).
Strictly speaking, the choice of eigenvalues is 
pseudorandom  (cf. Sec.~3.5 of Ref.~\cite{Knuth1998}) 
and the procedure 
is really only valid in the limit $N\to\infty$.
The details of this alternative extraction procedure are as follows.

Take $N=K\,n$ with positive integers $K$ and $n$. Then,
denote the sets of eigenvalues of the $n\times n$ blocks 
for the $M^1$ matrix by
\beqa\label{eq:calE1-all-blocks}
\mathcal{E}^1_\text{all\:blocks} &=&
\Big\{\mathcal{E}^1_1,\, \mathcal{E}^1_2,\,\ldots\,,\, \mathcal{E}^1_K \Big\}\,,
\\[2mm]
\mathcal{E}^1_k &=&
\Big\{\beta^1_{k,1},\, \beta^1_{k,2},\,\ldots\,,\, \beta^1_{k,n} \Big\}\,,
\eeqa
for $k \in \{1,\, 2,\,\ldots\,,\, K\}$, 
and similarly for the matrices $M^2$ and $M^3$
(the matrix $M^3$ is diagonal and the eigenvalues are
already known).
As said, the procedure is to take a pseudorandom element
of each set $\mathcal{E}^a_k$\,:
\beq\label{eq:xtilde-a-k}
\widetilde{x}^{\,a}_{k}=\beta^a_{k,\text{rand}[1,\,n]}/\sqrt{C_N}\,,
\eeq
where ``$\text{rand}[1,\,n]$'' is a uniform pseudorandom integer from the 
set $\{1,\, 2,\,\ldots\,,\, n\}$
and where a normalization factor has been included to
facilitate the comparison for different values of $N$.
The norm of each extracted spacetime point with coordinates
$\widetilde{x}^{\,a}_{k}$, for $a \in \{1,\, 2,\,3\}$
and $k \in \{1,\, 2,\,\ldots\,,\, K\}$, is defined by
\beq\label{eq:norm-xtilde-k}
|\widetilde{x}_{k}|   
\equiv 
\sqrt{
\left(\widetilde{x}^{\,1}_{k}\right)^2+
\left(\widetilde{x}^{\,2}_{k}\right)^2+
\left(\widetilde{x}^{\,3}_{k}\right)^2}\,.
\eeq

Let us clarify the procedure by considering
the explicit $9\times 9$ matrices from \eqref{eq:M3-M1-M2-N9}.
Taking three $3\times 3$ blocks along the diagonals
of these matrices then gives the following sets of eigenvalues:
\bsubeqs\label{eq:eigenval-N9-n3}
\beqa
\mathcal{E}^3 &=&
\Big\{ 
\big\{ -4,\, -3,\, -2 \big\}\,,
\big\{ -1,\, 0,\, 1   \big\}\,,
\big\{ 2,\, 3,\, 4    \big\}
\Big\}\,,
\\[2mm]
\mathcal{E}^1 &=&
\Big\{ \left\{ -{\sqrt{11/2}},{\sqrt{11/2}},0 \right\} ,
  \left\{ -{\sqrt{10}},{\sqrt{10}},0 \right\} ,
  \left\{ -{\sqrt{11/2}},{\sqrt{11/2}},0 \right\} 
\Big\}\,,
\\[2mm]
\mathcal{E}^2 &=&  \mathcal{E}^1\,.
\eeqa
\esubeqs
These results already show that the average eigenvalues vanish for
all blocks of $M^1$ and $M^2$, so that the averaging procedure of 
Sec.~\ref{sec:Extraction-procedure} becomes ineffective.
With a particular pseudorandom selection
in the alternative procedure \eqref{eq:xtilde-a-k}, 
we get from \eqref{eq:eigenval-N9-n3}
three points with the following Cartesian coordinates and
corresponding norms:
\bsubeqs\label{eq:xtildek-mum} %%numerics 28jan2021 14:55
\beqa
\widetilde{x}_{1} &=&
\left\{ 0,\, 0,\, -3/(2\sqrt{5}) \right\}\,,
\qquad\quad\quad\quad
|\widetilde{x}_{1}| =  3/(2\sqrt{5}) \approx 0.670820 \,,
\\[2mm]
\widetilde{x}_{2} &=&
\left\{ -1/\sqrt{2},\, -1/\sqrt{2},\, 0 \right\}\,,
\quad\quad\quad\;\;
|\widetilde{x}_{2}| = 1  \,,
\\[2mm]
\widetilde{x}_{3} &=&
\left\{ \sqrt{11/40},\, \sqrt{11/40},\, 1/\sqrt{5} \right\}\,,
\quad
|\widetilde{x}_{3}| = \sqrt{3}/2   \approx 0.866025\,.
\eeqa
\esubeqs

Results for larger values of $N$ are presented in 
Figs.~\ref{fig:fuzzy-sphere-N-128}--\ref{fig:fuzzy-sphere-N-555}.
In each of these four figures, two block sizes $n$ are considered 
and most extracted points are relatively
close to the unit sphere (see the respective bottom panels).
Also we can identify several ``strands'' along the meridians  
(see the respective three-dimensional views
of the northern and southern hemispheres), 
which are similar to the strands of Fig.~\ref{fig:fig3}
as discussed in Sec.~\ref{sec:Spacetime-points-from-test-2-master-field}.

The explanation of the strands here is especially  
straightforward and instructive. Consider 
Figs.~\ref{fig:fuzzy-sphere-N-128}--\ref{fig:fuzzy-sphere-N-512}
for $n=2$, which clearly shows four strands. 
In this case, the eigenvalues of the $2\times 2$ blocks 
of $M^1$ and $M^2$ are pairs 
of opposite numbers and the pseudorandom selection procedure
just picks the sign. For a fixed value of $|x^3|<1$, we then
have four symmetric points in the four quadrants
of the $(x^1,\,x^2)$ plane, which build the four strands as
the $x^3$ slice is varied.
For $n=4$ these four strands get somewhat ``scattered.''
The discussion is similar for the case of odd $n$ as shown in 
Figs.~\ref{fig:fuzzy-sphere-N-135}--\ref{fig:fuzzy-sphere-N-555}.

%%\newpage%%tmp
\subsection{General remarks}  
\label{subapp:General remarks}

The fuzzy-sphere matrices of Sec.~\ref{subapp:Fuzzy-sphere-matrices} 
have a perfectly regular structure 
(a dominant ``fine-structure'' in the terminology of
Secs.~\ref{sec:Test-master-fields} and \ref{sec:Discussion}), 
whereas the test matrices of 
Sec.~\ref{sec:Test-master-fields} have
a built-in scatter (randomness).
This explains that the averaging procedure of 
Sec.~\ref{sec:Extraction-procedure} 
gives nonzero results for the scattered test matrices of
Sec.~\ref{sec:Test-master-fields}
but not for the regular matrices of the fuzzy-sphere.

\begin{figure}[t]  %%[p]
\begin{center}
\hspace*{-0mm}
\includegraphics[width=1\linewidth]
{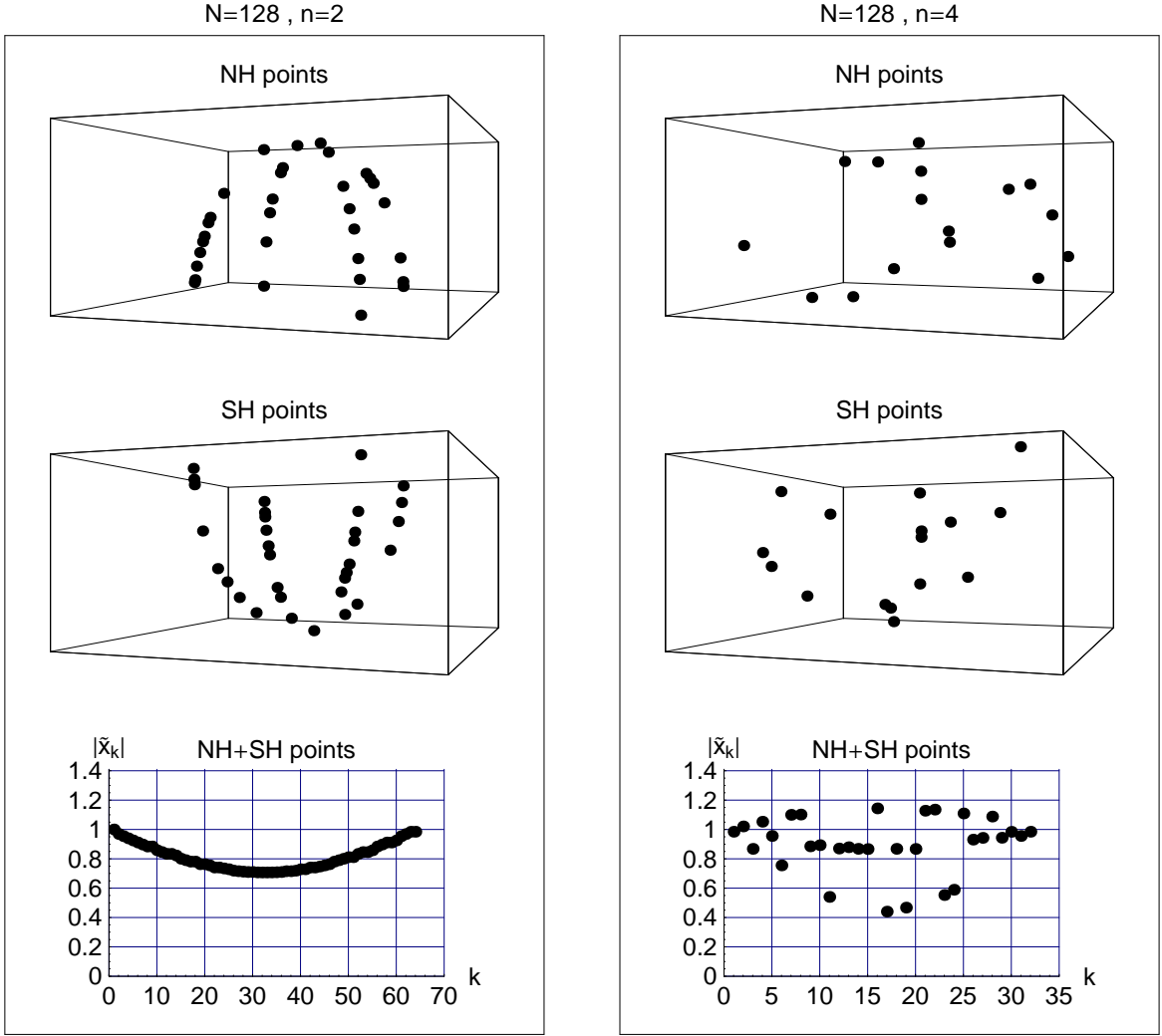}
%%{NCgeom-fuzzy-sphere-N128-n2-n4-v092.eps}
\end{center}
\caption{Fuzzy sphere $S_{2}^{N}$ for $N=2\,j+1=128$. 
The corresponding $N\times N$ matrices $M^a$ are defined
by \eqref{eq:fuzzy-sphere-matrices-M-def-N}%
--\eqref{eq:fuzzy-sphere-matrices-construction}.
An alternative extraction procedure, 
with block size $n=2$ (left) and $n=4$ (right),
gives emergent spacetime points with coordinates
$\widetilde{x}_{k}^{\:a}$, for $a \in \{1,\, 2,\,3\}$
and $k=1,\,2\, \ldots\, ,\, N/n$,
as defined by \eqref{eq:xtilde-a-k}.
Shown are the points of the northern hemisphere
(NH, $\widetilde{x}_{k}^{\:3}\geq 0$), 
the points of the southern hemisphere
(SH, $\widetilde{x}_{k}^{\:3}< 0$), and the
modulus $|\widetilde{x}_{k}|$ for all points
as defined by \eqref{eq:norm-xtilde-k}.
%The viewpoints for the northern and southern hemispheres 
%are slightly different.
}
\label{fig:fuzzy-sphere-N-128}
\vspace*{100mm}
\end{figure}

\begin{figure}[t]  %%[p]
\begin{center}
\hspace*{-0mm}
\includegraphics[width=1\linewidth]
{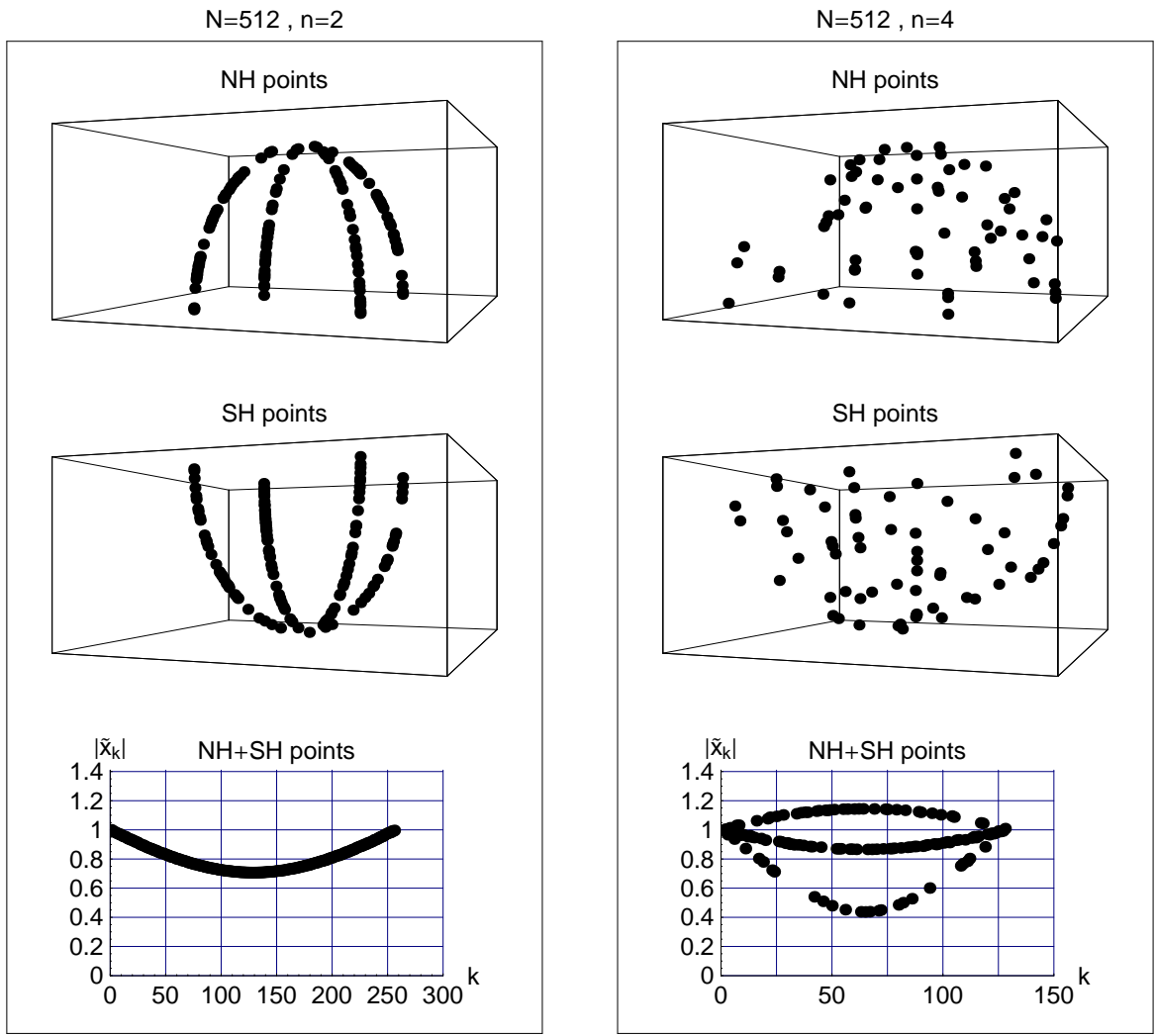}
%%{NCgeom-fuzzy-sphere-N512-n2-n4-v092.eps}
\end{center}
\caption{Same as Fig.~\ref{fig:fuzzy-sphere-N-128}, but now for
$N=512$.}
\label{fig:fuzzy-sphere-N-512}
\vspace*{100mm}
\end{figure}

\begin{figure}[t]  %%[p]
\begin{center}
\hspace*{-0mm}
\includegraphics[width=1\linewidth]
{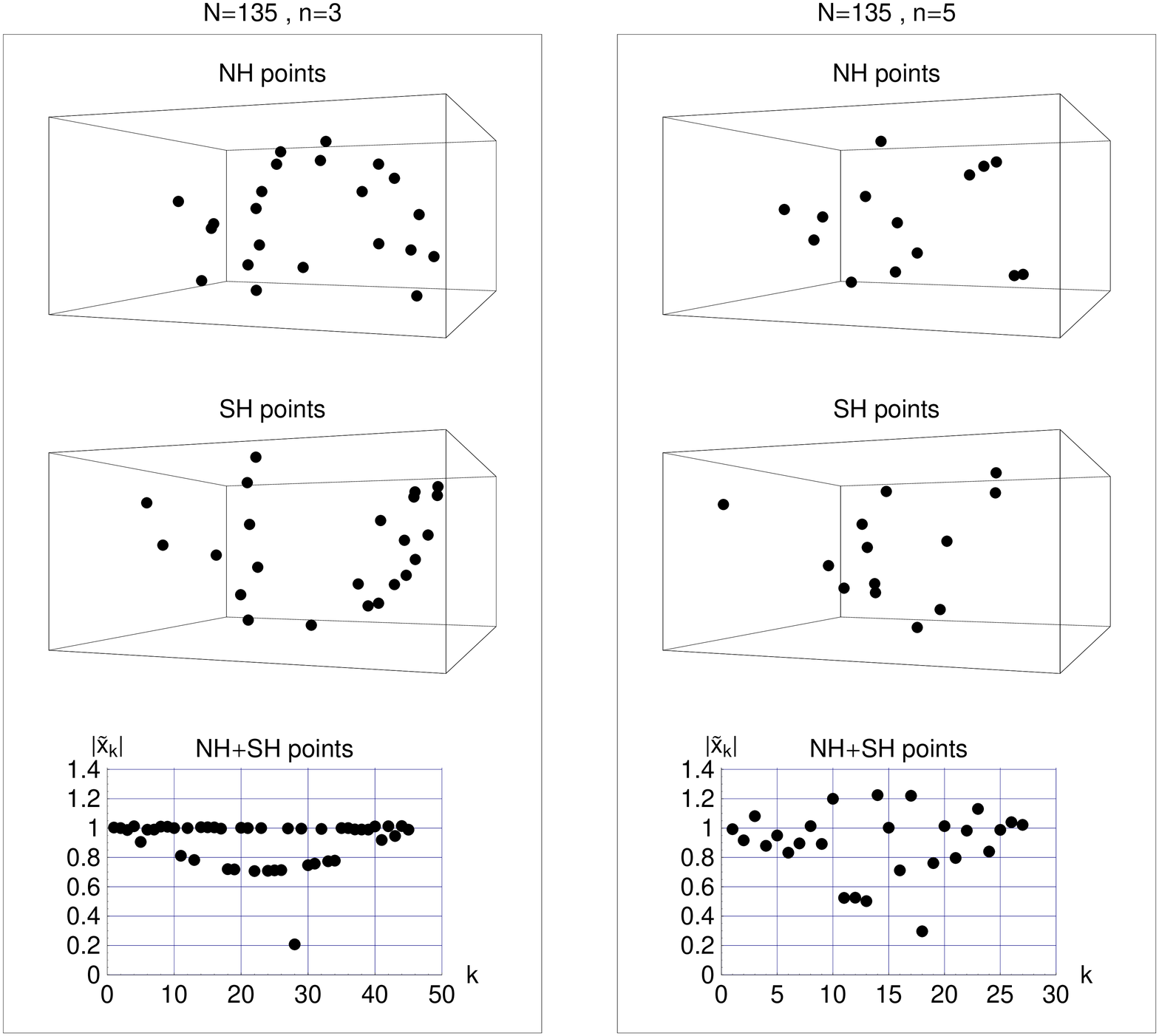}
%%{NCgeom-fuzzy-sphere-N135-n3-n5-v092.eps}
\end{center}
\caption{Same as Fig.~\ref{fig:fuzzy-sphere-N-128}, but now for
$N=135$ and $n=3,\,5$.}
\label{fig:fuzzy-sphere-N-135}
\vspace*{100mm}
\end{figure}

\begin{figure}[t]  %%[p]
\begin{center}
\hspace*{-0mm}
\includegraphics[width=1\linewidth]
{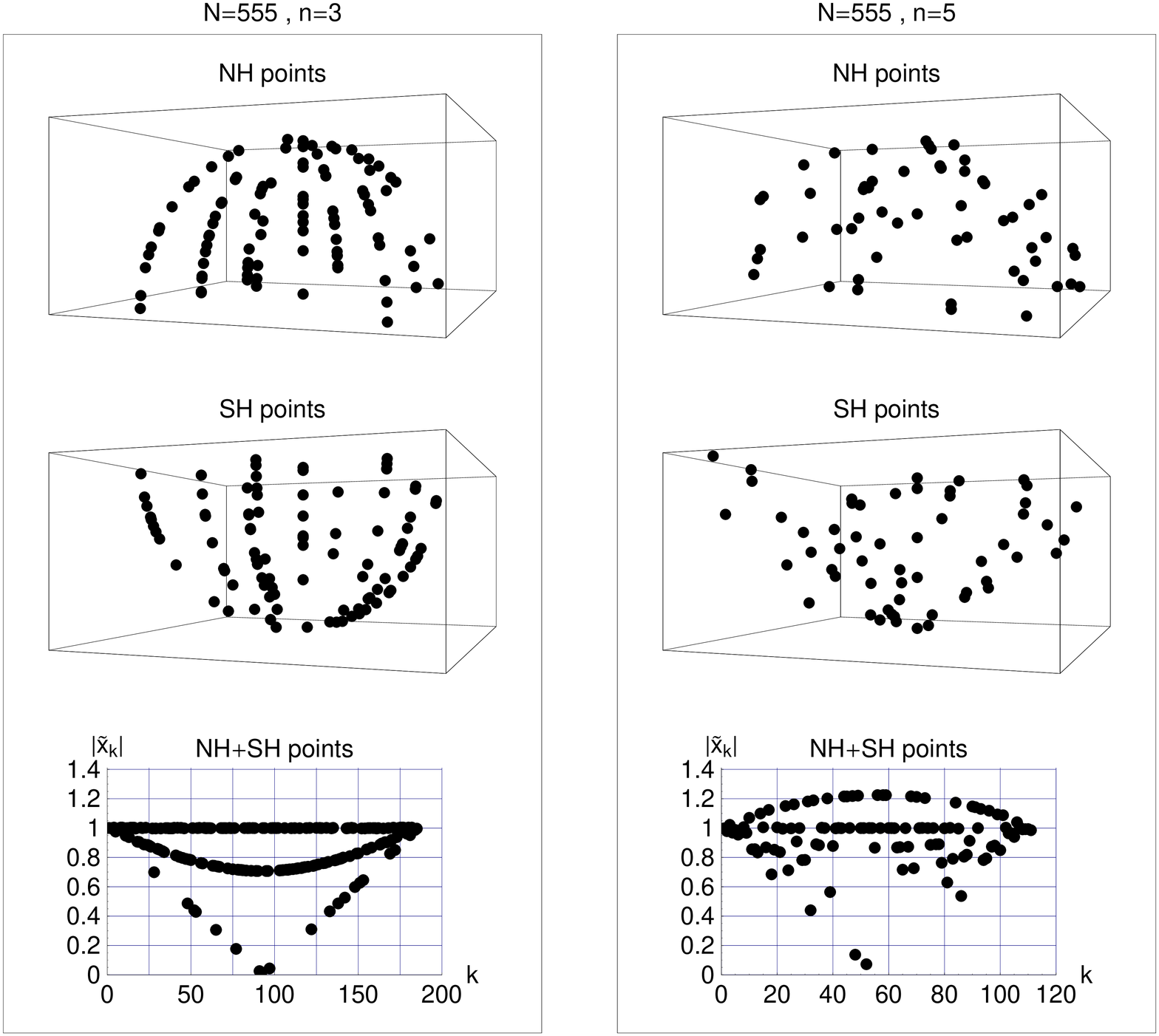}
%%{NCgeom-fuzzy-sphere-N555-n3-n5-v092.eps}
\end{center}
\caption{Same as Fig.~\ref{fig:fuzzy-sphere-N-128}, but now for
$N=555$ and $n=3,\,5$.}
\label{fig:fuzzy-sphere-N-555}
\vspace*{100mm}
\end{figure}

The alternative extraction procedure of 
Sec.~\ref{subapp:Alternative-extraction-procedure} carries
in itself a large degree of randomness, so that nonzero
results are obtained also from the regular fuzzy-sphere matrices 
(Figs.~\ref{fig:fuzzy-sphere-N-128}--\ref{fig:fuzzy-sphere-N-555}).
Most likely, the alternative procedure will work
equally well for the test matrices of Sec.~\ref{sec:Test-master-fields}, 
as long as the matrix dimension $N$ is large enough 
for a fixed value of block dimension $n$. 

A further issue is how well the $\text{O}(N/n)$ extracted points  
``fill-out'' an emergent space manifold, here the sphere $S^2$.
The four strands of the $n=2$ panels in 
Fig.~\ref{fig:fuzzy-sphere-N-512} are not very successful in
this respect, but the points of the $n=4$ panels in the same
figure are perhaps somewhat better. In the same way,
the $n=5$ panels in Fig.~\ref{fig:fuzzy-sphere-N-555}
may already cover the sphere reasonably well, apart from 
some ``outlier'' points with $|\widetilde{x}_{k}| \lesssim 0.7$.

In principle, it may be possible to obtain $\text{O}(N)$ points
if we consider \emph{overlapping} $n\times n$ blocks along the diagonals,
but still randomly select a single eigenvalue from each block
(overlapping blocks have also been considered 
in recent numerical work on the deformed Lorentzian IIB 
matrix model~\cite{NishimuraTsuchiya2019}).

\end{appendix}

\newpage
%\vspace*{-5mm}

\end{document}